\documentclass{aa}  
\usepackage{graphicx}
\usepackage{txfonts}
\begin{document}
            
\title{Effervescent heating: constraints from nearby cooling flow clusters observed with XMM-Newton}

\author{ R. Piffaretti \inst{1}
         \and
         J.S. Kaastra \inst{2}
         }

\offprints{R. Piffaretti}

\institute{Institut f\"ur Astrophysik, Leopold-Franzens Universit\"at
Innsbruck, Technikerstra\ss e 25, A-6020 Innsbruck, Austria
             \and
              SRON National Institute for Space Research,
              Sorbonnelaan 2, 3584 CA Utrecht, The Nether\-lands
          }

\date{Received  / Accepted  }

  \abstract
   {}
   {We have used deprojected radial density and temperature profiles 
of a sample
of 16 nearby CF clusters observed with \textit{XMM-Newton} to test whether the 
effervescent heating model can satisfactorily explain 
the dynamics of CF clusters.}
   {For each cluster we derived the required extra
heating as a function of cluster-centric distance for various values of the 
unknown parameters $\dot M$ (mass deposition rate) and $f_c$ (conduction
efficiency). We fitted the extra heating curve using the AGN effervescent 
heating function and derived the AGN parameters $L$ (the time-averaged
luminosity) and $r_0$ (the scale radius where the bubbles start rising in the
ICM).}
   {While we do not find any solution with the effervescent heating model 
for only one object, we do show that AGN and conduction heating are not 
cooperating effectively for half of the objects in our sample. 
For most of the clusters we find that, when a comparison is possible, the 
derived AGN scale radius $r_0$ and the observed AGN jet extension have the 
same order of magnitude. 
The AGN luminosities required to balance radiative 
losses are substantially lowered if the fact that the AGN deposits energy 
within a finite volume is taken into account. For the Virgo cluster, we find 
that the AGN power derived from the effervescent heating model is in good 
agreement with the observed jet power.}
  {}


\titlerunning{Effervescent heating in CF clusters}
\authorrunning{Piffaretti \& Kaastra}

\maketitle

\section{Introduction}
The radiative cooling time of the intracluster medium (ICM) in the
cores of many galaxy clusters is short enough for the plasma to
radiate an amount of energy equal to its thermal energy in less than a
billion years. Therefore the gas
should gradually condense, be replaced by the surrounding material, and
ultimately form a cooling flow (hereafter CF). Early estimates of the mass deposition
rates from X-ray imaging ($\sim 10^2 - 10^3 \mathrm{M}_{\odot}
\mathrm{yr}^{-1}$) have indicated 
that the cores of CF clusters should contain large amounts 
of cold gas, but searches for these large amounts of condensed baryons
 have not been successful (see Fabian \cite{fabian94} for a
 review of the subject of CFs before the launch of the {\it XMM-Newton} and
 {\it Chandra} satellites).

{\it XMM-Newton} and {\it Chandra} observations have greatly improved
 our understanding of CF clusters. Most important,
 X-ray spectra failed to detect the emission lines
that dominate the emission from gas below 2 keV 
(Peterson et al. \cite{peterson01}, Tamura et al. \cite{tamura01}, Kaastra et
 al. \cite{kaastra01}, Peterson et al. \cite{peterson03}), 
therefore ruling out the standard CF model. In addition, as
shown by Peterson et al. (\cite{peterson03}), the observed spectra cannot be explained
without any substantial modifications to the general CF
process. Recent observations have also shown that mass deposition rates are significantly smaller
than those estimated previously
(B{\" o}hringer et al. \cite{boehringer01}, David et al. \cite{david01},
 Peterson et al. \cite{peterson03}). Even though this implies a less
severe discrepancy between CF mass deposition and star-formation rates 
(McNamara \cite{mcnamara04}), no agreement between the amount of cold
baryons predicted from X-ray observations and the one observed in other
wavelengths has been found yet. 
Spatially-resolved spectroscopy
shows that the temperature in CF clusters drops towards the center to
approximatively one third of the cluster mean temperature, indicating
that the gas is prevented from cooling below these cutoff temperatures
(Peterson et al. \cite{peterson01}, Tamura et al. \cite{tamura01}, Kaastra et
 al. \cite{kaastra04}).

This recent evidence shows that the dynamics of the ICM in CF  
clusters is not solely governed by cooling of the ICM, and that some
heating and/or non X-ray cooling mechanisms must be
investigated. While non X-ray, rapid cooling mechanisms have been studied (see
Peterson et al. \cite{peterson03} and references therein), the most appealing
mechanisms involve heating processes.

Most CF clusters host a central active galactic
nucleus (AGN) with strong radio activity (Burns \cite{burns90}, Ball et
al. \cite{ball93}) and, 
most important, recent observations show that these
radio sources are interacting with the ICM and often displace the hot
gas, leaving cavities in their wakes (e.g., see Blanton \cite{blanton04}, B{\^
  i}rzan et al. \cite{birzan} and
references therein). Hence, it has been realized that gas heating by the
outflows of the central AGN can be a
vital process for the dynamics of CFs. 
This heating mechanism involves buoyant plasma bubbles inflated by
the AGN, which then rise through the cluster atmosphere, expand, and
ultimately heat the ICM (Churazov et al. \cite{churazov02}, Br{\" u}ggen \&
Kaiser \cite{bruggen01}, Ruszkowski \& Begelman \cite{ru02}, Br{\" u}ggen \&
Kaiser \cite{bruggen02}). 
Although there is no consensus regarding the ability of AGN heating to
  prevent the formation of CFs and the efficiency of the processes by 
which energy is delivered to the ICM, there is evidence suggesting that 
the class of models in which the AGN energy input alone balances 
radiative losses is unable to quench the CF (Brighenti \& Mathews
\cite{brighenti02}, Zakamska \& Narayan \cite{zakamska03}).

Despite the fact that the presence of magnetic fields in clusters implies
that thermal conductivity is only a fraction of the Spitzer value, 
thermal conduction by electrons might play an important role in CFs. 
While some theoretical work has pointed out that thermal conduction must be
highly suppressed (e.g., see Fabian \cite{fabian94}), recent
theoretical papers show that conductivity can be as high as a
substantial fraction of the Spitzer rate (Narayan \& Medvedev
\cite{narayan01}, Gruzinov \cite{gruzinov02}) and
therefore it has been reconsidered as a possible heat source candidate. 
However, it has been shown that heat conduction alone fails to balance radiative losses
at the center of the clusters (Voigt et al. \cite{voigt02}, Zakamska \&
Narayan \cite{zakamska03}, Voigt \&
Fabian \cite{voigt04}, Kaastra et al. \cite{kaastra04}) and that models with
heat conduction as the only heating source are unstable
(Soker \cite{soker03}).

Other, less popular heating mechanisms are: fluid turbulence generated
by substructure motion or cluster mergers (Fujita et al. \cite{fujita04}), heating
from intra-cluster supernovae (Domainko et al. \cite{domainko04}), heating by hadronic
cosmic rays (Colafrancesco et al. \cite{colafrancesco04}), and heating by magnetic fields
(Soker \& Sarazin \cite{soker90}, Makishima et al. \cite{makishima01}).

The inability of heat conduction to balance radiative losses and the
  failure of some AGN heating models have motivated the development of 
models in which these two mechanisms cooperate. Models with heat conduction and AGN heating acting
together are very attractive due to the complementary nature of the
two processes: thermal conduction is effective/ineffective in the outer/inner regions
of the cluster, and AGN heating is effective/ineffective in the
inner/outer part. Ruszkowski \& Begelman (\cite{ru02}) (hereafter RB02) have recently proposed
such a model. The ICM density and temperature evolved according to
their model reach a final stable configuration in agreement with the general
shape of observed density and temperature profiles in CF clusters. RB02's 
model (or the \textit{effervescent heating} model, hereafter) is the only 
proposed theoretical model
that can be effectively tested against observations. Unfortunately this has
only been done for the Virgo cluster (Ghizzardi et al. \cite{ghizz}), and it is not
clear whether the model can give a satisfactory explanation of the dynamics of
CF clusters in general.

In the present work we address the latter question using a sample of CF
clusters observed with {\it XMM-Newton}. Our sample, which consists of 16
objects, is large for this kind of analysis and all the objects were 
analyzed using an appropriate and homogeneous procedure.

The layout of this paper is as follows. In Sect.~\ref{sect:data} we
present the sample and briefly summarize the procedure used to derive
the basic quantities used in our analysis. In Sect.~\ref{sect:equations} we give a description
of RB02's model and present the equations used in
Sect.~\ref{sect:results}, where we fit the model to the data. In
Sect.~\ref{sect:conclusions} we summarize the results of our analysis and
present the conclusions.

We adopt $H_{\mathrm{0}} =70$~km\,s$^{-1}$\,Mpc$^{-1}$,
$\Omega_{\mathrm{m}}=0.3$ and $\Omega_{\mathrm{\Lambda}}=0.7$ in all
the calculations throughout this paper.
\section{Data\label{sect:data}}
The sample consists of the clusters listed in
Table~\ref{tab:tfitstable}. These objects are part of the sample analyzed
in Kaastra et al. (\cite{kaastra04}; hereafter K04), who derived deprojected
radial temperature and density profiles for all the clusters in the
sample, including a description of the sample and an extensive
presentation of the data analysis. From the K04's sample, we excluded 4
clusters: Coma, A~754 and A~3266, because these are non-CF clusters, and A~1837
since temperature and density could be measured in only one bin within the cooling radius. 
Spectral fitting was done over the full 0.2--10~keV range and, in general, 
temperatures and gas densities were computed for the innermost 8
 shells (see K04, Table 3 for the boundaries
 between the shells), except when the data became too noisy in the
outermost shell. The sample used in the present work therefore consists of
16 clusters with CF for which gas density and temperature radial profiles are well measured. Since
thermal conduction depends on temperature gradients and the AGN
heating in the RB02 model involves a dependency
on ICM pressure, we model gas temperature and density profiles by
fitting the data with analytic functions. We used more that one
fitting function for both quantities in order to quantify the
difference due to the modelling. In the following we describe these
functions and the fitting to the data.   

\subsection{Modelling of gas temperature and density \label{sect:dataTmod}}
The deprojected temperature profiles of our clusters exhibit
a self-similar shape studied in Piffaretti et al. (\cite{piff05}): the temperature declines
from the maximum cluster temperature at a break radius
$r_{\mathrm{br}}$ moving outwards and shows the characteristic
temperature decline towards the X-ray emission peak. Since we are
interested in the central cooling region and the cooling
radius for a cooling time of 15 Gyr, $R_{\mathrm{cool}}$ is smaller than
$r_{\mathrm{br}}$ for all the clusters, the temperature
profiles can be simply modelled by a function that is monotonically
raising with radius. Hence, for each cluster we select temperature
bins inside the radius $R_{\mathrm{T,max}}=r_{\mathrm{br}}$ and fit them using the following expressions:
\begin{equation}\label{eqn:tfitA}
T(r) = T_0 + T_1 \frac{(r/r_\mathrm{T})^{\mu}}
{1+(r/r_\mathrm{T})^{\mu}},
\end{equation}
or 
\begin{equation}\label{eqn:tfitE}
T(r) = \tilde{T}_0 - \tilde{T}_1 \, \mathrm{exp} \Big( - \frac{r^2}{2 \tilde{r}^2_{T}}\Big) \,.
\end{equation}
In order to reduce the number of parameters here, we set $T(r=0)$ equal
to the temperature of the central bin for both fits and use $\mu=2$ in
Eq.~\ref{eqn:tfitA} (Allen et al. \cite{allen01}). For Perseus, Virgo, A~262,
A~496, and A~3112, the temperature in the observed radial range is monotonically
raising with radius, and therefore $R_{\mathrm{T,max}}$
is just the distance of the outermost bin from the cluster center. The
best-fit parameters and $R_{\mathrm{T,max}}$ are given in
Table~\ref{tab:tfitstable}. Both temperature parametrizations are used in the
computation of thermal conduction and in the modelling of the gas
pressure. Our main results presented in Sect.~\ref{sect:results} are
achieved using the parametrization in Eq.~\ref{eqn:tfitA}, and we use the second
parametrization given in Eq.~\ref{eqn:tfitE} only in order to explore the effect of
a different modelling on our main results. The changes introduced by using
Eq.~\ref{eqn:tfitE} are, as we will show in Sect.~\ref{sect:modelinfl}, quite
small and do not change the results obtained by using Eq.~\ref{eqn:tfitA}.

\begin{table*}[!ht]
\caption{\label{tab:tfitstable} The radius at the center of the last radial bin
considered in the temperature fits and the best-fit parameters from
fitting Eqs.~\ref{eqn:tfitA} and ~\ref{eqn:tfitE} to the data ($1 \, \sigma$ errors
in parenthesis). Here, $\mu=2$ in Eq.~\ref{eqn:tfitA} is used and, for both
fitting functions, $T(r=0)$ is set equal to the temperature of the central bin.}
\centerline{
\begin{tabular}{|l|c|ccc|ccc|}
\hline
Cluster               &$R_{\mathrm{T,max}}$&$T_0$&$T_1$&$r_{\mathrm{T}}$&$\tilde{T}_0$&$\tilde{T}_1$&$\tilde{r}_{\mathrm{T}}$\\
                      &(kpc)&(keV)&(keV)&(kpc)&(keV)&(keV)&(kpc)\\
\hline
NGC 533               &75&0.67(0.01)&0.81(0.03)&18(1)&1.35(0.17)&0.68(0.16)&17(4)\\
Virgo                 &35&1.45(0.02)&1.27(0.18)&16(3)&2.45(0.13)&1.00(0.11)&11(2)\\
A 262                 &199&1.02(0.02)&1.24(0.08)&21(2)&2.16(0.08)&1.14(0.05)&16(1)\\
S\'ersic~159$-$3      &233&2.17(0.05)&0.29(0.04)&38(12)&2.42(0.07)&0.25(0.02)&30(5)\\
MKW 9                 &167&1.28(0.09)&1.40(0.31)&57(16)&2.42(0.24)&1.14(0.15)&40(6)\\
2A 0335+096           &206&1.40(0.03)&1.76(0.16)&52(6)&2.88(0.14)&1.48(0.11)&38(3)\\
MKW 3s                &268&3.00(0.10)&0.57(0.12)&40(20)&3.52(0.21)&0.52(0.10)&36(14)\\
A 2052                &212&1.41(0.06)&1.79(0.14)&27(5)&3.03(0.15)&1.62(0.09)&22(2)\\
A 4059                &275&2.11(0.13)&2.17(0.15)&44(7)&4.03(0.23)&1.92(0.09)&35(3)\\
Hydra A               &321&2.92(0.10)&0.63(0.12)&114(30)&3.46(0.19)&0.54(0.08)&88(16)\\
A 496                 &289&2.14(0.10)&2.35(0.20)&59(7)&4.20(0.22)&2.06(0.12)&45(3)\\
A 3112                &645&2.99(0.08)&1.55(0.21)&61(17)&4.36(0.20)&1.37(0.12)&46(8)\\
A 1795                &369&3.47(0.09)&2.67(0.20)&90(9)&5.73(0.19)&2.26(0.10)&65(4)\\
A 399                 &404&2.60(0.57)&4.29(0.44)&67(17)&6.61(0.91)&4.01(0.34)&58(11)\\
Perseus               &229&3.07(0.05)&3.91(1.10)&132(28)&5.76(0.58)&2.69(0.53)&79(11)\\
A 1835                &594&5.13(0.14)&2.78(0.24)&207(68)&7.34(0.77)&2.21(0.63)&134(50)\\
\hline\noalign{\smallskip}  
\end{tabular}
}
\end{table*}
We model the gas density by using a single $\beta$-model given by: 
\begin{equation}
\rho(r)=\rho_{0} \Big(1+(r/r_c)^2 \Big)^{-\frac{3}{2} \beta}\,. 
\label{eq:sbeta}
\end{equation}
The density profile is fitted within $R_{\mathrm{\rho,max}}$, which
is the radius at the center of the last radial bin, where a robust estimate 
of gas density and temperature is possible 
(see Piffaretti et al. \cite{piff05} for the bin selection criterion). 
The best-fit parameters for the single $\beta$-model and
$R_{\mathrm{\rho,max}}$ are given in Table ~\ref{tab:sbetatable}.

An alternative parametrization of the gas density profile is the more complex
double $\beta$-model, which is a popular generalization of the single
$\beta$-model used to model the central surface-brightness excess
observed in CF clusters. Unfortunately in this case, the gas density is
modelled using the sum of two single $\beta$-models, so the number of 
free parameters is doubled: $\rho_{0,i}$, $\beta_i$, and
$r_{c,i}$, with $i=1,2$. As a consequence, while fitting the single
$\beta$-model to the density profiles gives statistically significant values 
for the best-fit parameters,
the large number of parameters adopted in the double $\beta$-model, 
together with the small number of bins in which the gas density is
measured, do not allow a significant determination of the best-fit
parameters. Therefore, we decided to present our main results in
Sect.~\ref{sect:results} using the single $\beta$-model. 
Nonetheless, in order to constrain to which extent the double $\beta$-modelling
changes the results, we also fit the density profiles using a double $\beta$-model
with a reduced number of fitting parameters, by setting 
$\beta_1=\beta_2$ and fixing one of the core radii equal to
the core radius $r_T$ (the core radius $\tilde{r}_T$ is
also used as a supplementary choice). 
The derived profiles are then used to model the gas pressure as done
by using the single $\beta$-model results. A discussion of the influence of this
different modelling on our main results is given in
Sect.~\ref{sect:modelinfl} below. 
\begin{table}[!ht]
\caption{\label{tab:sbetatable} The radius at the center of the last
radial bin
considered in the gas density fits and the best-fit parameters from
fitting Eq.~\ref{eq:sbeta} to the data ($1 \, \sigma$ errors
in parenthesis).}
\centerline{
\begin{tabular}{|l|c|ccc|}
\hline
Cluster               &$R_{\mathrm{\rho,max}}$&$\rho_0$&$r_{\mathrm{c}}$&$\beta$\\
                      &(kpc)&($10^{-20} \, \mathrm{g}/\mathrm{m}^3$)&(kpc)&\\
\hline
NGC 533               &160&1.85(0.77)&7(2)&0.46(0.01)\\
Virgo                 &35&8.14(0.85)&3(1)&0.33(0.01)\\
A 262                 &199&4.54(0.91)&5(1)&0.33(0.01)\\
S\'ersic~159$-$3      &499&4.36(0.06)&35(1)&0.59(0.01)\\
MKW 9                 &238&67.7(16.29)&19(8)&0.32(0.02)\\
2A 0335+096           &309&8.24(0.49)&20(2)&0.53(0.01)\\
MKW 3s                &402&2.61(0.12)&35(2)&0.49(0.01)\\
A 2052                &319&4.02(0.36)&19(2)&0.45(0.02)\\
A 4059                &412&1.71(0.18)&48(7)&0.50(0.02)\\
Hydra A               &673&5.13(0.70)&29(5)&0.52(0.02)\\
A 496                 &193&5.14(0.14)&16(1)&0.42(0.01)\\
A 3112                &645&4.15(0.35)&41(4)&0.55(0.02)\\
A 1795                &553&4.26(0.18)&42(3)&0.52(0.01)\\
A 399                 &606&0.72(0.08)&76(15)&0.38(0.03)\\
Perseus               &229&7.27(0.52)&28(3)&0.53(0.02)\\
A 1835                &831&5.24(0.23)&80(5)&0.62(0.02)\\
\hline\noalign{\smallskip}  
\end{tabular}
}
\end{table}
\section{Effervescent heating\label{sect:equations}}
Assuming spherical symmetry, the thermodynamic equations describing
the ICM can be written in the form:
\begin{eqnarray}
\label{eq:eqs1}
&& v\rho r^2 = \mathrm{const} = {{\dot M } \over {4 \pi}}\\
\label{eq:eqs2}
&& M_{tot} = - {{r^2 v }\over {G}} {{dv} \over{dr}} -  {{kTr} \over {G \mu m_p}}
\left[ {{d \ln{T} } \over {d \ln{r}}}
 +  {{d\ln{\rho}} \over {d \ln{r}}} \right] \\
\label{eq:eqs3}
&& H =\varepsilon - \varepsilon_{cond}+\varepsilon^\star 
\end{eqnarray}
where
\begin{equation}
\varepsilon^\star =   {{\rho v k T } \over {\mu m_p r}}
\left[ \frac 32 {{d\ln{T}} \over{d\ln{r}}} -  {{d\ln{\rho}} \over{d \ln{r}}} \right] \, ,
\label{eq:epsilonstar}
\end{equation}
and $v$ is the gas flow velocity which is taken positive outwards and $\dot M $ is
the mass flow rate, which is therefore negative for an
inflow and assumed, throughout the paper, to be constant. Here, $M_{tot}$ is the total
gravitational mass within the radius $r$ and $\rho$, $T$ are the gas density
and temperature, respectively ($\mu=0.61$ is
the mean molecular weight). Further, $\varepsilon=n_{\mathrm{e}}^2 \Lambda
(T)$ is the plasma emissivity and $\varepsilon^\star$ is the energy due to the
inflow/outflow of the gas. Then, $\varepsilon$ is computed from the deprojected
  electron density $n_{\mathrm{e}}$ and $\Lambda (T)$, the cooling 
function for a plasma with an average metallicity $Z=0.5 \, Z_{\odot}$ losing energy
by bremsstrahlung and line emission. The quantity $H$ in the energy equation
(Eq.~\ref{eq:eqs3}) is thus an extra heating term. In the framework of the
effervescent heating model
developed in RB02, the extra heating is provided by buoyant bubbles injected into the ICM by the
central AGN (see this section below).

The total gravitational mass $M_{tot}$ given in Eq.~\ref{eq:eqs2} differs from the
mass estimated under the assumption of hydrostatic equilibrium, since
it includes a velocity term. For the values of $\dot M$ considered
in Sect.~\ref{sect:results} below, this term is negligible, so the mass 
analysis presented in Piffaretti et al. (\cite{piff05}) gives accurate
values for the mass profiles of the clusters in the present work.

For all the clusters in our sample, the entropy of the gas increases
monotonically moving outwards, almost proportional to the radius
(Piffaretti et al. \cite{piff05}). Given this evidence, the convection term in the
original effervescent heating model is neglected, since a declining
entropy profile is essential for convection to operate (see the discussion
in Ghizzardi et al. \cite{ghizz}). The term $\varepsilon_{cond}$ is the
heating due to thermal conduction, which is given by
\begin{equation}
\varepsilon_{cond} = 
\frac 1{r^2} {d \over {dr}} \left( r^2  \kappa {{dT} \over {dr}}\right) , 
\label{eq:epscond}
\end{equation}
where $\kappa$ is the conductivity. For an ionized plasma,
$\kappa$ is the Spitzer conductivity:
\begin{equation}
\kappa=\kappa_S = {{1.84 \times 10^{-5} \left(T\right)^{5/2} } \over 
{\ln{\Lambda}}} 
{\rm \, erg \, cm^{-1}\, s^{-1}\, K^{-1}} \, ,
\label{eq:kappa}
\end{equation}
where $\ln{\Lambda} \sim 40$ is the Coulomb logarithm.

Magnetic fields at the $\mu$G level are known to be present in 
clusters of galaxies, while higher values, up to tens of $\mathrm{\mu}$G, have been measured at
the center of CF clusters (see Govoni \& Feretti \cite{govoni04} for a recent
review). 
In the presence of
such fields thermal conduction is suppressed below the Spitzer rate
given in Eq.~\ref{eq:kappa} by a factor $\sim 100-1000$ (Binney \& Cowie
\cite{binney81}, Chandran \& Cowley \cite{chandran98}). This condition has been essential for the
development of multiphase CF models (e.g., Fabian \cite{fabian94}), but
recent work has shown that the level of suppression of thermal
conduction might not be as high as previously thought. In particular,
Narayan \& Medvedev (\cite{narayan01}) have shown that if the magnetic field is chaotic over
a wide range of length scales, thermal conduction is enhanced to $\sim
1/5$ times the Spitzer value. Gruzinov (\cite{gruzinov02}) pointed out that the
effective heat conduction in a random variable magnetic field is
boosted to a factor of 3 below the Spitzer value. The latter
result yields the upper limit for the efficiency of heat conduction in
the ICM. Therefore, in our analysis presented in
Sect.~\ref{sect:results}, we investigate models with thermal
conduction varying from zero to the maximum level of $1/3$ times the
Spitzer value. The ICM conductivity $\kappa$ is therefore characterized by the
fraction $f_c$ according to $\kappa=f_c \times \kappa_S$.

Since thermal conduction is most efficient at high temperatures and it generally
fails to balance radiative losses in the central parts of CF clusters, a
heat source able to supply energy in the central parts is needed. 
Such a central heat source is incorporated in the RB02 model, in which AGNs are assumed to inject buoyant bubbles into
the ICM, which heat the ambient medium by doing $p\, d\,V$ work as
they rise and expand adiabatically. Here we summarize the derivation of the 
AGN heating function $H^{AGN}$ (Begelman \cite{begelman01}). 
Assuming a {\it steady state}, the energy flux avaliable for heating is:
\begin{equation}
\dot e \propto p_{b}^{(\gamma_b -1 )/\gamma_b},
\label{eq:enrgyflux}
\end{equation}
where $p_{b}$ is the pressure of the bubbles and $\gamma_{b}$ their
adiabatic index. Assuming that the bubbles' partial pressure scales as
the ICM pressure $p$, the heating function $H^{AGN}$ can be expressed according to:
\begin{equation}
 H^{AGN} = - h(r) \left({p \over p_0}\right)^{(\gamma_b-1)/\gamma_b} \frac 1r 
{{d\ln{p}}\over {d\ln{r}}}, 
\label{eq:H}
\end{equation}
with 
\begin{equation}
h(r)= {L \over {4 \pi r^2}} \left( 1 - e^{-r/r_0} \right) q^{-1}
\label{eq:h}
\end{equation}
and where
\begin{equation}
q=\int_{0}^{+\infty}{\left({p \over {p_0}}\right)^{(\gamma_b-1)/\gamma_b}
\frac 1r {{d\ln{p}}\over {d\ln{r}}} \left( 1 - e^{-r/r_0} \right) dr} \, .
\label{eq:q}
\end{equation}
Here, $p_0$ is some reference ICM pressure (in the following we
take its value at the cluster center) and $L$ the
{\it time-averaged} luminosity of the central source. 
The term $ 1 - \mathrm{exp}(-r/r_0)$ introduces an inner cutoff that 
fixes the scale radius where the bubbles start rising in the ICM. 
Therefore an inner cutoff is already taken into account when
performing the integration in Eq.~\ref{eq:q} and with zero as the lower
integration limit gives a correct estimate for the {\it time-averaged}
luminosity of the AGN. It must be noted, however, that the upper integration limit in Eq.~\ref{eq:q} should be
replaced by a finite number, i.e. the radius within which the bubbles
effectively deposit heat into the ICM. This issue is
explored in Sect.~\ref{sect:int} since it is through Eq.~\ref{eq:q}
that the {\it time-averaged} luminosity of the central source is
estimated and a finite integration limit has the effect of diminishing
its energy requirement.

Finally, we notice that the following results are not
only valid for the AGN-heating mechanism just described (i.e., heating by
buoyant bubbles), but for any mechanism for which its dependence on the ICM
pressure is equivalent to the one given in Eq.~\ref{eq:H}. 
\section{Results\label{sect:results}}
From the observed deprojected gas temperature and density profiles, the required extra
heating $H$ can be computed for fixed values of the mass
deposition rate $\dot M$ and the conduction efficiency $f_c$, and then fitted using the
AGN heating function $H^{AGN}$. For each bin with measured gas
density and temperature, we compute the gas emissivity $\varepsilon$. For a
fixed pair $(\dot M,f_c)$, the conductive heat $\varepsilon_{cond}$ and the heat term $\varepsilon^\star$ are
computed using the temperature and gas parametrizations given in    
Eqs.~\ref{eqn:tfitA} and ~\ref{eq:sbeta}, since they depend on the gradients
of these quantities. Then $\varepsilon_{cond}$ and $\varepsilon^\star$ are 
evaluated at the radius where the gas
emissivity is computed, to finally obtain extra heating $H$ through
Eq.~\ref{eq:eqs3}.

The errors of $H$ are computed from the errors ($1 \, \sigma$) of the gas
emissivity and the $\varepsilon_{cond}$ and $\varepsilon^\star$ errors. The
latter have been evaluated by propagating the $1 \, \sigma$ errors of the gas
density and temperature best-fit parameters. While two different radial ranges have been used to fit the
gas density and temperature profiles ($R_{\mathrm{\rho,max}}$ and
$R_{\mathrm{T,max}}$, see Sect.~\ref{sect:dataTmod}), the extra heating $H$ is
computed in bins within $R_{\mathrm{T,max}}$, since $R_{\mathrm{T,max}} <
R_{\mathrm{\rho,max}}$. Finally the
AGN heating function given in Eq.~\ref{eq:H} is fitted to the extra heating
data points using a $\chi^2$ minimization. The gas pressure and pressure
gradients in Eq.~\ref{eq:H} are evaluated using Eqs.~\ref{eqn:tfitA} and 
~\ref{eq:sbeta}. For any fixed pair $(\dot M,f_c)$ the free
parameters are therefore $r_0$ and the luminosity of the central AGN $L$. 
The bubbles' adiabatic index $\gamma_{b}$
is fixed to $4/3$ throughout the paper (i.e., we assume relativistic bubbles), since the inclusion of $\gamma_{b}$ as an
additional free parameter leads to very large errors in the derived
best-fit parameters. A discussion of the effects due to the use of
different gas density and temperature parametrizations is given in
Sect.~\ref{sect:modelinfl}.
\subsection{Zero AGN heating\label{sect:standard0}}
Despite the stability issues mentioned above, the case of thermal conduction 
by electrons acting as the only heat
source has been already explored quite extensively in the literature,
since it provides immediate information on the strength of heating by
conduction in different clusters. 
Starting from Eqs.~\ref{eq:eqs1}~-~\ref{eq:eqs3} one can estimate
whether heat conduction alone can be efficient in quenching the
cooling flow by setting both $H$ and $\dot M$ equal to zero. For
a sample that includes all the objects studied in the present work, 
K04 have explored this possibility and concluded, in agreement with
other studies (Voigt et al. \cite{voigt02}, Voigt \& Fabian \cite{voigt04}), that heat conduction
alone is insufficient to balance radiative losses in CF clusters (with
the exception of MKW~9 and A~399, see also discussion below), in
particular in their central regions. These results and the growing
observational evidence of AGN-ICM interaction at the center of CF clusters
motivate including in the model an additional heating term
 provided by a central AGN.
\subsection{Conduction and AGN heating: zero mass-dropout \label{sect:M0}}
Motivated by the results of RB02, who find that the effervescent
heating model does not require any mass dropout, we first present
results for $\dot M=0$. As this case is the most justified, we present
 its predictions in more detail. In addition, as we show below, small
 mass-inflow/outflow rates do not modify the
results substantially. For $\dot M=0$, we vary
$f_c$ between $0$ and $1/3$ and fit the extra heating curve $H$
(Eq.~\ref{eq:eqs3}) using the heating function $H^{AGN}$ (Eq.~\ref{eq:H}). 
No constraints are imposed on the AGN
luminosity $L$, but the best-fit values for $r_0$ are searched in the
interval from 0 to the cooling radius $R_{\mathrm{cool}}$, which
is computed for a cooling time of 15 Gyrs and is listed in
Table~\ref{tab:M0res}.

\begin{figure}
\resizebox{\hsize}{!}{\includegraphics{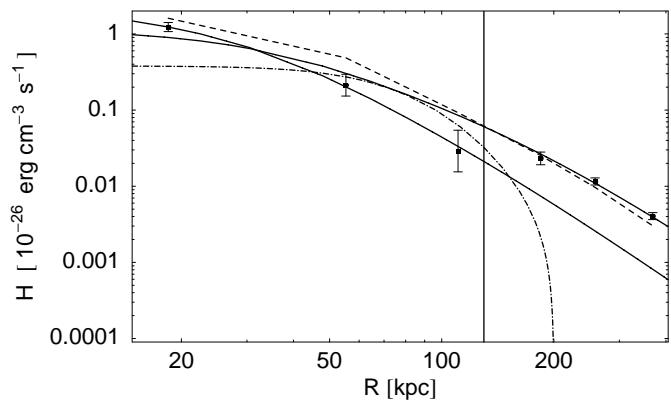}}
\caption{Energy requirements in A~1795. The plasma emissivity is shown,
    for simplicity, by the dashed straight lines joining the emissivity values
    computed in bins and heating due to thermal conduction by the dot-dashed
    line. The mass flow rate is set to zero in this model, so the extra 
heating curve (filled squares) is determined by $f_{\mathrm{c}}
\times$ Spitzer. Here $f_{\mathrm{c}}=1/3$, the maximum value allowed
in the models. The three bins inside the cooling radius (vertical line) 
alone are best fitted by an inner cutoff radius
$r_0=14$ kpc and {\it time-averaged} AGN luminosity $L=8.0 \times 10^{44}
\, \mathrm{erg} \, \mathrm{s}^{-1}$. For the same model ($\dot{M}=0$ and 
$f_{\mathrm{c}}=1/3$), all six bins are
best fitted by an unphysical $r_0=179$ kpc and very high AGN
luminosity $L=3.3 \times 10^{45}\, \mathrm{erg} \, \mathrm{s}^{-1}$. The two 
solid curves are the best-fit functions for the two cases. See text for
  details.}    
\label{fig:a1795}
\end{figure}

In Fig.~\ref{fig:a1795} we illustrate results for the cluster A~1795
and discuss them in the following, since the model outcome for this object
highlights important features also found for most of the clusters in the sample. Since no
mass dropout is present, $\varepsilon^\star$ is zero in Eq.~\ref{eq:eqs3}
and the extra heating $H=\varepsilon - \varepsilon_{\mathrm{cond}}$ only
depends on the conduction efficiency $f_c$. Since $f_{\mathrm{c}}=1/3$ 
is the maximum value we consider, it corresponds to the maximum 
energy yield by heat conduction from the
outer parts of the cluster. From a visual inspection of
Fig.~\ref{fig:a1795}, it is clear that heat conduction is not able to
lower the extra heating in the outermost bins. As a consequence, if
the extra heating curve is fitted with Eq.~\ref{eq:H}, the resulting
best-fit parameter $r_\mathrm{0}$ (the scale radius where the bubbles start
rising in the ICM) is unphysically large ($r_0=176$ kpc $>\,
R_{\mathrm{cool}}$, see Table~\ref{tab:M0res}) and the {\it
time-averaged} AGN luminosity $L=3.3 \times 10^{45} \, \mathrm{erg} \,
\mathrm{s}^{-1}$ is also quite large. 
The latter solution is not taken into account since, as stated above,
 the best-fit value for $r_0$ is searched in the
interval from 0 to $R_{\mathrm{cool}}$, but is used here as an
example. 

On the other hand, if the effervescent heating is assumed to be efficient in the region
only within the cooling radius, one can notice that heat conduction is
effective in reducing the extra heating so that the resulting extra
heating curve is monotonically falling with radius. In this case one therefore
expects that the extra heating supplied by the raising relativistic
bubbles must be distributed over smaller distances and that the total
AGN energy output is lower. This is indeed reflected in the much
different best-fit parameters $r_0=14$ kpc and $L=8 \times 10^{44}
\, \mathrm{erg} \, \mathrm{s}^{-1}$. This feature indicates that, as expected, the effervescent heating
model strongly depends on how much and at which radial distance heat
conduction is efficient. Because of this, we performed, for each
cluster and each analytic function used to model the gas density and
temperature, fits using the whole observed radial range (i.e., $0< r < R_{\mathrm{T,max}}$) and the radial
range delimited by the cooling radius.

Another common feature is the
effect of the variation of the conduction efficiency. If
$f_{\mathrm{c}}=0$ the extra heating curve is simply equal to the gas
emissivity and the increase in $f_{\mathrm{c}}$ from 0 to $1/3$ gives
a decrease in the extra heating curve from the emissivity curve to the
data points marked in Fig.~\ref{fig:a1795}. The
increase in $f_{\mathrm{c}}$ should hence lead to a decrease in the
AGN energy requirement. In fact we find that, within the cooling
radius, both {\it time-averaged} AGN luminosity and inner cutoff radius
$r_0$ decrease monotonically with increasing $f_{\mathrm{c}}$. 
This opposite effect is seen if the fits are performed over the whole
radial range: an increase in $f_{\mathrm{c}}$ leads to an increase in
both $L$ and $r_0$, again showing the inadequacy of applying the model
over the whole radial range.

For the majority of the objects in our sample, we find that 
the effervescent heating
model provides results when the radial range used to fit the extra
heating curve with Eq.~\ref{eq:H} is not simply the whole observed range 
but the radial range inside the cooling region. In fact for
NGC~533, Virgo, A~262, MKW~9, A~4059, and A~496, only fits performed
over the radial range inside the cooling region converge. The reason for this
lack of convergence of the fits is, as in the case of A~1795 illustrated in
Fig.~\ref{fig:a1795}, that heat conduction is not able to
reduce the extra heating in the outermost bins. On the other hand, a reduction
of the extra heating is present in bins close to the cooling radius. 

This is a welcome feature for the effervescent heating model, 
because it implies that heating from the central
source is not only more viable in terms of total energy output, but also that
the energy can be distributed from the center outwards up to relatively small
distances. For S\'ersic~159$-$3, 2A~0335+096, Hydra~A, and A~1835 heat conduction is
low even for the maximal conduction efficiency $f_c=1/3$. 
Therefore a characteristic drop of the extra heating close to
the cooling radius is not present in these objects and hence
no difference is seen between the best fits performed using different
radial ranges. The lowering of the extra heating close to
the cooling radius is also not seen in MKW~3s, A~2052, and A~3112,
because heat conduction is quite efficient only at the cluster
center. While in A~3112 heat conduction is low and therefore the best-fit 
parameters for the two radial ranges are almost identical, for
MKW~3s and A~2052 we obtain slightly higher values for both $r_0$ and $L$ when
the whole radial range is used instead of the cooling region.
 
For these reasons, only results from fits to the extra heating 
curves within the cooling
radius will be discussed. For A~399, no results are obtained for either of the
two radial ranges. The reason for the lack of convergence of the fits is that
in A~399 heat conduction is very efficient, owing to its relatively high
temperature and very steep temperature profile (see also K04, Table 11). 
Conduction can balance radiative losses completely for $f_c$ as low as 0.03. 
For lower values of the conduction efficiency the best-fit values for 
$r_0$ are always larger than the cooling radius $R_{\mathrm{cool}}=43 \,
\mathrm{kpc}$ and therefore not taken into account. 

The results for the zero mass dropout models are summarized in
Table~\ref{tab:M0res}. 
For each value of $f_c$ the 1 $\sigma$ errors on the best-fit
parameters are very small compared to the range spanned by the best-fit 
parameters for the different models and therefore are not
reported. In addition, it is important to notice that when fits
are restricted to the cooling region, the number of bins we are using
in the fits is quite low (3 for 9 objects, 4 for 4 objects, and 5 for 3
objects), and although it is sufficient for our scope,
it should be noticed that for more complex models (e.g., models with
$\gamma_{b}$ as an additional free parameter) the number of bins
within the cooling radius should be increased if possible. While for most of the clusters we
find a solution for $L$ and $r_0$ for every value of $0<f_c<1/3$, for
some objects the fits converge only for models with $f_c$ in a narrower 
interval
($f_{\mathrm{c}}^{\mathrm{min}}$-$f_{\mathrm{c}}^{\mathrm{max}}$, see
Table~\ref{tab:M0res}). 
In particular, in clusters where conduction is high (MKW~9
and Perseus), solutions are found for conduction efficiencies 
substantially lower than the maximum allowed value $f_c=1/3$. In 
S\'ersic~159$-$3, 2A~0335+096, Hydra~A, A~3112, and A~1835, the conductivity
yield is low and, as expected, different models (i.e., different values of
$f_c$) give almost identical results.
 
It is crucial to highlight the trend of the best-fit parameters of the different models with model parameter $f_c$. For most of
the clusters, the model with $f_{\mathrm{c}}=f_{\mathrm{c}}^{\mathrm{max}}$ is
the one where values for both $L$ and $r_0$ are lowest 
and for $f_{\mathrm{c}}=f_{\mathrm{c}}^{\mathrm{min}}$ the highest 
($L^{\mathrm{min,max}}$ and $r_{\mathrm{0}}^{\mathrm{min,max}}$ 
in Table~\ref{tab:M0res}). In addition $L$ and $r_0$ vary monotonically with $f_c$ within these
limits. The clusters that exhibit this trend are labeled by an asterisk in
column {\it trend} of Table~\ref{tab:M0res}. For MKW~3s, A~2052, and
A~3112 the trend just described is reversed. In these objects heat conduction 
lowers the extra heating curve especially in the cluster center. This implies
more and more flattening of the extra heating curve at the center with
increasing $f_c$, which is finally reflected in the increase in both $r_0$ and
$L$. In Table~\ref{tab:M0res} we also list the best-fit parameters $\tilde{L}$
and $\tilde{r}_0$ for the best
model ($\tilde{f}_c$), which is chosen by simply comparing the $\chi^2$ of the best fit for the
different models.
\begin{table*}[!ht]
\caption{\label{tab:M0res} Cooling radius $R_{\mathrm{cool}}$ and the results for the
effervescent heating model with no mass dropout ($\dot{M}=0$). For each model,
i.e. for a fixed conduction efficiency $0<f_{\mathrm{c}}< 1/3$, the extra
heating curve $H$ derived from Eqs.~\ref{eq:eqs1}~-~\ref{eq:eqs3} is
fitted using Eq. ~\ref{eq:H} over the radial range $0<r<R_{\mathrm{cool}}$, and
the best-fit values for $L$ and $r_0$ are derived. Only models with
$f_{\mathrm{c}}$ between $f_{\mathrm{c}}^{\mathrm{min}}$ and
$f_{\mathrm{c}}^{\mathrm{max}}$ can be fitted. 
The corresponding minimum and maximum values of the {\it time-averaged} luminosity $L$
and the inner cutoff radius $r_0$ are given. For most of the clusters, both
best-fit parameters for {\it time-averaged} luminosity and
inner cutoff radius decrease monotonically with increasing
$f_{\mathrm{c}}$. These are labeled by an asterisk in the column {\it trend}.
$\tilde{L}$ and $\tilde{r}_0$ are the {\it time-averaged}
luminosity and the inner cutoff radius for the
model with the smallest reduced $\chi^2$ and $\tilde{f}_c$ is the corresponding conduction
efficiency. The minimum
and maximum values of the {\it time-averaged} luminosity
$L^{\mathrm{min,max}}_{\mathrm{fin}}$ for finite integration limits (see
Sect.~\ref{sect:int}) are also listed. No result is found for A~399.}
\centerline{
\begin{tabular}{|l|c|cccccc|c|ccc|}
\hline
Cluster
&$R_{\mathrm{cool}}$&$f_{\mathrm{c}}^{\mathrm{min}}$&$f_{\mathrm{c}}^{\mathrm{max}}$&$r_{\mathrm{0}}^{\mathrm{min}}$&$r_{\mathrm{0}}^{\mathrm{max}}$&($L^{\mathrm{min}}$,
$L^{\mathrm{min}}_{\mathrm{fin}}$)&($L^{\mathrm{max}}$, $L^{\mathrm{max}}_{\mathrm{fin}}$)&{\it
  trend}&$\tilde{r}_{\mathrm{0}}$&$\tilde{L}$&$\tilde{f}_{\mathrm{c}}$\\
   &(kpc)& & & (kpc)& (kpc)&
   ($\mathrm{erg} \, \mathrm{s}^{-1}$)& ($\mathrm{erg} \, \mathrm{s}^{-1}$)& &(kpc)&($\mathrm{erg} \, \mathrm{s}^{-1}$)& \\
\hline
NGC 533               &36&0.00&0.26&0.3&1.5&$(1.6, 0.5) \times 10^{42}$&$(2.6,0.8) \times 10^{42}$&*&1.5&$2.6   \times 10^{42}$&0.00\\
Virgo                 &24&0.03&0.26& 0.6& 17.9&$(3.8,1.5)  \times 10^{42}$ & $(2.1,0.4) \times 10^{43}$&*&0.9&$4.0   \times 10^{42}$&0.24\\
A 262                 &61&0.12&0.33&6.6&60.4&$(8.1,2.5)  \times 10^{42}$&$(4.7,0.6)  \times 10^{43}$&*&6.6&$8.1  \times 10^{42}$&0.33\\
S\'ersic~159$-$3      &128&0.00&0.33&21.6&22.6&$(5.8,2.2)  \times 10^{44}$& $(5.8,2.3) \times 10^{44}$&*&22.6&$5.8 \times 10^{44}$&0.00\\
MKW 9                 &74&0.08&0.10&14.9&14.9&$(5.7,0.5)  \times 10^{42}$&$(6.5,0.6)  \times 10^{42}$&*&14.9&$5.7  \times 10^{42}$&0.10\\
2A 0335+096           &121&0.00&0.33&14.6&16.1&$(6.1,2.3)  \times 10^{44}$&$(6.8,2.6)  \times 10^{44}$&*&14.6&$6.1  \times 10^{44}$&0.33\\
MKW 3s                &95&0.00&0.33&37.2&94.0&$(43.3,0.2)  \times 10^{43}$&$(58.1,7.3)  \times 10^{43}$&&65.9&$5.0  \times 10^{44}$&0.23\\
A 2052                &114&0.00&0.33&6.6&21.1&$(2.6,0.7)  \times 10^{44}$&$(3.4,0.8)  \times 10^{44}$&&17.6&$3.4  \times 10^{44}$&0.25\\
A 4059                &86&0.00&0.33&4.7&53.8&$(7.4,1.5)  \times 10^{43}$&$(5.0,0.6)  \times 10^{44}$&*&4.7&$7.4  \times 10^{43}$&0.33\\
Hydra A               &130&0.00&0.33&26.5&29.4&$(7.0,2.6) \times 10^{44}$&$(7.7,2.8)  \times 10^{44}$&*&26.5&$7.0  \times 10^{44}$&0.33\\
A 496                 &104&0.07&0.28&11.1&86.3&$(1.2,0.4)  \times 10^{44}$&$(5.7,0.8)  \times 10^{44}$&*&86.3&$5.7  \times 10^{44}$&0.07\\
A 3112                &141&0.00&0.33&29.9&58.7&$(1.3,0.3)  \times 10^{45}$&$(1.4,0.4)  \times 10^{45}$&&58.7&$1.4  \times 10^{45}$&0.33\\
A 1795                &130&0.00&0.33&13.5&48.5&$(8.0,2.3)  \times 10^{44}$&$(2.4,0.5) \times 10^{45}$&*&19.2&$1.0 \times 10^{45}$&0.28\\
A 399                 &43&-&-&-&-&-&-&-&-&-&\\
Perseus               &128&0.01&0.18&40.0&73.4&$(1.6,0.5)  \times 10^{45}$&$(2.5,0.6)  \times 10^{45}$&*&40.0&$1.6  \times 10^{45}$&0.18\\
A 1835                &204&0.00&0.33&33.2&37.8&$(1.0,0.3)  \times 10^{46}$& $(1.1,0.4)  \times 10^{46}$&*&33.2&$1.0  \times 10^{46}$&0.33\\
\hline\noalign{\smallskip}  
\end{tabular}
}
\end{table*}
\begin{table*}[!ht]
\caption{\label{tab:Mnot0res} Results for the
effervescent heating model with mass dropout or outflow. The
minimum and maximum values of the {\it time-averaged} luminosity $L$ and the inner cutoff radius $r_0$ are computed from the results of
models with conduction efficiency $0<f_c< 1/3$ and
$-\dot{M}^{\mathrm{max}}/10<\dot{M}< + \dot{M}^{\mathrm{max}}/10$. The minimum
and maximum values of the {\it time-averaged} luminosity
$L^{\mathrm{min,max}}_{\mathrm{fin}}$ for finite integration limits (see Sect.~\ref{sect:int}) are also listed. As
for the $\dot{M}=0$ models, no result is found for A~399.}
\centerline{
\begin{tabular}{|l|c|cccc|}
\hline
Cluster               &
$\dot M^{\mathrm{max}}$
&$r_{\mathrm{0}}^{\mathrm{min}}$&$r_{\mathrm{0}}^{\mathrm{max}}$&($L^{\mathrm{min}}$,
$L^{\mathrm{min}}_{\mathrm{fin}}$)&($L^{\mathrm{max}}$, $L^{\mathrm{max}}_{\mathrm{fin}}$) \\
                      &($M_{\odot}/yr$)&(kpc)&(kpc)&($\mathrm{erg} \, \mathrm{s}^{-1}$)&($\mathrm{erg} \, \mathrm{s}^{-1}$)\\
\hline
NGC 533               &5&0.2&13.5&$(1.4,0.3)  \times 10^{42}$&$(3.1,1.0)   \times 10^{42}$\\
Virgo                 &6&0.2&31.4&$(3.3,1.3)  \times 10^{42}$&$(3.4,0.5)  \times 10^{43}$\\
A 262                 &10&5.4 &60.5 &$(6.1,2.0)  \times 10^{42}$&$(5.4,0.7)  \times 10^{43}$\\
S\'ersic~159$-$3      &210&20.9&23.4&$(5.5,2.1)  \times 10^{44}$&$(6.1,2.4) \times 10^{44}$\\
MKW 9                 &11&10.9&38.5 &$(4.1,0.4)  \times 10^{42}$&$(16.2,0.9)  \times 10^{42}$\\
2A 0335+096           &420& 14.4&16.1 &$(5.2,2.0)  \times 10^{44}$&$(7.7,2.9)  \times 10^{44}$\\
MKW 3s                &45& 35.3&94.4 &$(4.3,0.5)  \times 10^{44}$&$(5.8,0.6)  \times 10^{44}$\\
A 2052                &100&4.6 &89.4 &$(2.5,0.6)  \times 10^{44}$&$(4.5,0.9)  \times 10^{44}$\\
A 4059                &100&2.6 &67.3 &$(5.2,1.0)  \times 10^{43}$&$(5.1,0.7)  \times 10^{44}$\\
Hydra A               &180&26.0 &30.1 &$(6.6,2.4)  \times 10^{44}$&$(8.1,2.9)  \times 10^{44}$\\
A 496                 &120&10.0 &89.3 &$(1.1,0.3)  \times 10^{44}$&$(6.9,0.7)  \times 10^{44}$\\
A 3112                &300&27.7 &72.2 &$(1.3,0.3)  \times 10^{45}$&$(1.5,0.4)  \times 10^{45}$\\
A 1795                &380&12.8 &52.5 &$(6.6,1.9)  \times 10^{44}$&$(2.5,0.5) \times 10^{45}$\\
A 399                 &50&-&-&-&-\\
Perseus               &650&34.4&78.5 &$(1.5,0.4)  \times 10^{45}$&$(2.8,0.6)  \times 10^{45}$\\
A 1835                &5800&24.1 &53.6 &$(6.5,2.2)  \times 10^{45}$&$(1.5,0.4)  \times 10^{46}$\\
\hline\noalign{\smallskip}  
\end{tabular}
}
\end{table*}
\subsection{Conduction and AGN heating: models with mass dropout or outflow \label{sect:Mnot0}}
Although RB02's results indicate that an equilibrium state with no mass
dropout is usually achieved, the effervescent heating model
incorporates the possibility of gas inflow or outflow. It is therefore
interesting to see how much our results are changed by including 
gas inflow/outflow (negative/positive mass flow rates), since this
implies a positive/negative work (per unit volume) done by the
system ($\varepsilon^\star$ in Eq.~\ref{eq:eqs3}). It is expected that 
negative mass-flow rates reduce the energy
required by the central source to quench cooling, but the question is
how large this effect is.

We apply the effervescent heating model as done in Sect.~\ref{sect:M0}, but
we go on to explore models with
$-\dot{M}^{\mathrm{max}}/10<\dot{M}< +
\dot{M}^{\mathrm{max}}/10$. When possible, the maximum mass flow rate
$\dot M^{\mathrm{max}}$ is set equal to the maximum of the mass deposition 
rates given in Table 5 in
Peterson et al. (\cite{peterson03}). Otherwise $\dot M^{\mathrm{max}}$ is set equal to the
maximum value that, at $f_{\mathrm{c}}=0$, still gives positive values for
$H$ inside the cooling radius. The values of the maximum mass flow rates
$\dot M^{\mathrm{max}}$ are listed in Table~\ref{tab:Mnot0res}. This choice of
the upper and lower bounds for $\dot{M}$ is dictated by the fact that we find 
the effect of a fixed mass inflow/outflow rate for all the clusters 
to be drastically different from cluster to cluster. In addition, taking
RB02's results into account, low mass flow rates should be present when 
an equilibrium state (which is assumed here) is achieved. 
The analysis of models with $-\dot{M}^{\mathrm{max}}/10<\dot{M}< +
\dot{M}^{\mathrm{max}}/10$ is therefore a good compromise. We do not present 
the results from fitting the models extensively. 
More important we find that, as expected, the effect
introduced by mass inflow or outflow is to reduce the {\it
time-averaged} luminosity $L$ for inflows and enhancing it for
outflows. From the models we have studied and, of course, for those where the 
fits converge in the $L-r_0$ plane, we report minimum and maximum values
for the {\it time-averaged} luminosity $L$ and the inner
cutoff radius $r_0$ in Table~\ref{tab:Mnot0res}. A comparison with the values
for the model with no mass dropout (Table~\ref{tab:M0res}) shows that the 
broadening of the range for the best-fit parameters is not large.     
\subsection{Effects of the temperature and density modelling\label{sect:modelinfl}}
In order to investigate the effects introduced by the choice of
fitting functions for the density and temperature profiles, we 
performed the same analysis presented up to here (i.e., using Eq.~\ref{eqn:tfitA} and
Eq.~\ref{eq:sbeta}) but using the remaining possible combinations of
parametrizations given in Sect.~\ref{sect:dataTmod}. We find that for a fixed
model (i.e., for fixed $\dot M$ and $f_c$) the use of
different parametrizations introduces small differences in the
estimated best-fit parameters. These are smallest for the models
without mass dropout or outflow and in particular for those clusters in
which heat conduction is inefficient. For
S\'ersic~159$-$3, 2A~0335+096, Hydra~A, and A~1835, the difference in
both $L$ and $r_0$ is less than $3 \%$ and for the rest of the
clusters less than $10 \%$. The differences are
largest when the double $\beta$-model and Eq.~\ref{eqn:tfitE} are used
to parametrize gas density and temperature for models with
$\dot{M}\not=0$. Also in this case, the difference is smallest for the
clusters in which heat conduction is not efficient (typically
less then $7 \%$ for both best-fit parameters $L$ and $r_0$). For the
rest of the clusters, the difference can be as high as $18 \%$, but
generally less than $14 \%$. Taking other parametrizations of the 
temperature profile like those used in Churazov et al. (\cite{churazov03}) and
Dennis \& Chandran (\cite{dennis05}) 
leads to a change of at most $10 \%$ for both best-fit parameters $L$ and
$r_0$ for a fixed model. Most important, the trends between the AGN parameters 
$L$ and $r_0$ and the model parameters $f_c$ and $\dot{M}$ described in Sects.~\ref{sect:M0}
and~\ref{sect:Mnot0} remain unchanged for every combination of gas density and temperature
parametrizations used. We therefore conclude that the values given in Tables
~\ref{tab:M0res} and~\ref{tab:Mnot0res} give quite robust estimates
for the {\it time-averaged} luminosity $L$ and the inner cutoff radius $r_0$.
\subsection{Effects of the integration limits\label{sect:int}}
Since most of the models in which radiative cooling is balanced by some heating source
fail due to insufficient energy yield, it is important to investigate 
to what extent our values for the {\it time-averaged} luminosity of
the central AGN are overestimated. 
When fitting the extra heating function (Eq.~\ref{eq:H}) to the extra
heating curve, one directly derives the best-fit values for $r_0$ and
the normalization of the extra heating function, i.e. $L /4 \pi
q$. The {\it time-averaged} luminosity $L$ is then estimated using
Eq.~\ref{eq:q}. The values of $L$ reported so far were computed
by setting the integration limits as given in Eq.~\ref{eq:q}, but as
already noted in Sect.~\ref{sect:equations}, the upper integration limit
should be replaced. As reported in
Sect.~\ref{sect:M0}, the effervescent heating
model provides results for almost all the clusters 
when applied to the cooling region. As a
consequence we set the upper integration limit in in Eq.~\ref{eq:q}
equal to the cooling radius $R_{\mathrm{cool}}$. 

The effect of this replacement is to reduce the {\it time-averaged} luminosity of the
central AGN. The effect of this simple but effective correction is shown in
Table~\ref{tab:M0res}, where we
list the minimum and maximum {\it time-averaged} luminosities
$L^{\mathrm{min}}_{\mathrm{fin}}$ and
$L^{\mathrm{max}}_{\mathrm{fin}}$ computed using this finite
integration limit for the models with no mass dropout. The same quantities
for the model with mass dropout are listed in Table~\ref{tab:Mnot0res}. 
Comparing these values with the ones computed in the previous sections
shows that the reduction in the {\it time-averaged} luminosity can be 
almost one order of magnitude, hence a finite integration limit in
Eq.~\ref{eq:q} should be used when estimating the effervescent heating 
best-fit parameters.
\section{Discussion and conclusions\label{sect:conclusions}}
We have used deprojected radial density and temperature profiles of a sample
of 16 nearby CF clusters observed with {\it XMM-Newton} (Kaastra et al. \cite{kaastra04}) to
test whether the effervescent heating model (Ruszkowski \& Begelman
\cite{ru02}) can satisfactorily explain the structure of CF clusters. 
The effervescent heating model incorporates both heat conduction and AGN
feedback as heating
sources, which is of great interest because of the complementary nature of
these two processes. For each cluster, we derived the required extra
heating as a function of cluster-centric distance for various values 
of the unknown parameters $\dot M$ (mass deposition rate) and $f_c$ 
(conduction efficiency). We fitted the extra heating curve using 
the AGN-heating function proposed by Ruszkowski \&
Begelman (\cite{ru02}) and
derived the AGN parameters $L$ (the {\it time-averaged} luminosity) and $r_0$ 
(the scale radius where the bubbles start rising in the ICM).

For models without gas mass dropout or outflow ($\dot{M}=0$), we find:
\begin{itemize}
\item for only one object (A~399) we do not find any solution for 
the effervescent heating model because heat conduction is very efficient;  
\item for 4 clusters (S\'ersic~159$-$3, 2A~0335+096, Hydra~A, and A~1835) the
  conductivity yield is extremely low and, as expected, different models (i.e., different
  values of $f_c$) give almost identical results;
\item for 3 objects (MKW~3s, A~2052, and A~3112), we find that heat conduction
  plays an important role only at the cluster center and that, as a
  consequence, the trend between the fitted AGN parameters and conduction
  efficiency is not the one expected if conduction and AGN heating
  are assumed to cooperate;
\item for the remaining 8 clusters (NGC~533, Virgo, A~262, MKW~9, A~4059,
  A~496, A~1795, and Perseus), conduction and AGN heating are found to be
  cooperating.  
\end{itemize}
We have studied models with mild gas mass dropout of outflow
($-\dot{M}^{\mathrm{max}}/10<\dot{M}< + \dot{M}^{\mathrm{max}}/10$, with
$\dot{M}^{\mathrm{max}}$ being the mass deposition rate required by the
standard CF model) and find that:
\begin{itemize}
\item the conclusions listed above for the case of no mass dropout/outflow
  are also valid in this case;
\item as expected, the effect introduced by mass inflow or outflow is to 
reduce the required AGN heating for inflows and to enhance it for outflows;
\item this implies that the ranges of allowed AGN parameters $L$ and $r_0$ is
  broader than in the case $\dot{M}=0$, but we show that the broadening is not large. 
\end{itemize}
We find that our results are not sensitive to the choice of fitting
functions used to model gas density and temperature profiles and that the
{\it time-averaged} AGN luminosities required to balance radiative losses are
substantially reduced if the fact that the AGN deposits energy within a finite
volume is taken into account.

Since we do not find any solution with the effervescent heating model 
for only one object (A~399), we conclude that the model 
provides a satisfactory explanation of the observed structure of CF
clusters. As pointed out by Begelman (\cite{begelman04}), AGN-heating 
is dominant in the final stable state of the cluster modelled by Ruszkowski \&
Begelman (\cite{ru02}). Conductive heating dominates only at the beginning of
the evolution and might be essential only for the stability of the model. 
Therefore, models with high heat conduction are not preferred. 

On the other hand, since there is evidence that AGN heating alone 
is not able to quench CFs
(Brighenti \& Mathews \cite{brighenti02}, Zakamska \& Narayan
\cite{zakamska03}), it is fair to assume that thermal conduction,
although operating at different rates from cluster to cluster, must play an
important role as a heating mechanism. 
While this is found for 8 clusters in our
sample (NGC~533, Virgo, A~262, MKW~9, A~4059, A~496, A~1795, and Perseus), we
have shown that conductive heating is either completely unimportant in 
4 clusters
(S\'ersic~159$-$3, 2A~0335+096, Hydra~A, and A~1835), too high for one object 
(A~399), or high enough to play an important role but peaked at the cluster 
center in 3 clusters (MKW~3s, A~2052, and A~3112). 
Therefore, if we assume that AGN and conduction heating must be cooperating 
effectively, the model does not provide a satisfactory explanation for 
half of the objects in the sample.

These findings prompt us to posit that, at least for these
objects, the description of their thermal structure through a {\it steady
  state} solution of the thermodynamic equations is not viable and that we are
observing them at an evolutionary stage far from equilibrium. A clearer
picture can, of course, be achieved by studying a much larger sample using the
procedure employed in this work.

For the clusters that are well described by the effervescent heating model,
the derived best fit parameters $r_0$ and $L$ can be compared with results
obtained from complementary observations. In the effervescent heating scenario, it is supposed that bubbles are 
deposited by the central radio source. Although the nature of this process is 
not well understood, an interesting possibility is that bubbles are 
generated through the interaction of the radio jets with the surrounding ICM. 
In the framework of the effervescent heating model, the AGN parameter 
$r_0$ fixes the scale radius where the bubbles start rising in the cluster 
atmosphere and heat the ICM. One then expects the 
AGN parameter $r_0$ to be larger than the jet extension and of the same 
order of magnitude. Since the observed jet extension $r_{\mathrm{jet}}$ is at
most as large as the true jet size due to projection effects, the same
relationship is expected between $r_0$ and $r_{\mathrm{jet}}$. 

\begin{figure}
\resizebox{\hsize}{!}{\includegraphics{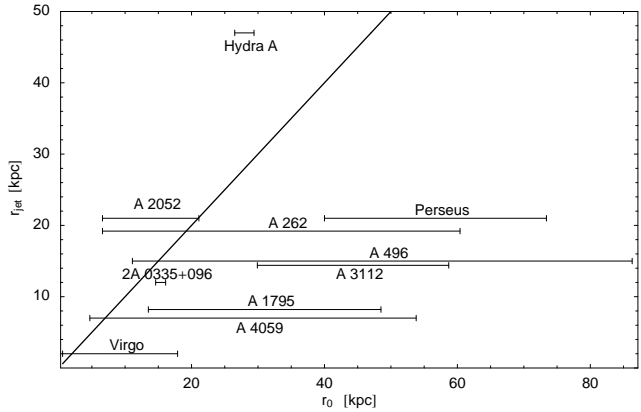}}
\caption{Comparison between the observed extension of the AGN jet
  $r_{\mathrm{jet}}$ and the best-fit AGN parameter $r_0$. For $r_0$ we plot
  the whole range of allowed values for the models with no mass dropout (see
  Table~\ref{tab:M0res}), and the extension of the AGN jet has been taken from
  the literature: Virgo (Young et al. \cite{young02}), A~262 (Blanton et
  al. \cite{blanton2004}), 2A 0335+096 (Sarazin et al. \cite{sarazin95}),
  A~2052 (Blanton et al. \cite{blanton03}), A~4059 (Taylor et
  al. \cite{taylor94}), Hydra A (McNamara et al. \cite{mcnamara00}), A~496 (Markovi{\'c} et
  al. \cite{markovic04}), A~3112 (Takizawa et al. \cite{takizawa03}), A~1795 (Ettori et al. \cite{ettori02}), and Perseus 
(Fabian et al. \cite{fabian02}). 
The solid line indicates the locus of equality.}\label{fig:r0rjet}
\end{figure}

The size of the AGN jet has only been measured for 10 of the 15 clusters 
that are well described by the effervescent heating model. 
For these 10 objects, we show the comparison between $r_0$ and 
$r_{\mathrm{jet}}$ in Fig.~\ref{fig:r0rjet}. 
Even though the range of allowed values for $r_0$ is quite 
large, the comparison shows that, for most of the clusters, the extension 
of the AGN jet and the scale radius $r_0$ are within the same order of 
magnitude and that the latter is in general larger than the jet extension. 
This implies that the possibility of bubbles being generated through 
the interaction of the radio jets with the ICM cannot be excluded.  

A comparison between the derived luminosities $L$ with
observed luminosities is by far more difficult. In fact, in the framework
of the effervescent heating model, the derived AGN luminosity is a
time-averaged {\it total} AGN power. Since the radio power is a poor
tracer of the total power (Eilek \cite{eilek04}) a fair comparison is possible
only for objects with estimated total jet power. At present, estimating 
the total jet power is unfortunately possible for only one
source: M~87 in the Virgo cluster (Owen et al. \cite{owen00}). In this case, our estimate
of $L$ (see Table~\ref{tab:M0res}) agrees well with the inferred
total jet power, $\sim 3-4 \times 10^{42}\,\, \mathrm{erg} \mathrm{s}^{-1}$. A
similar conclusion is presented in Ghizzardi et al. (\cite{ghizz}).
       
Even though we used the largest sample of CF clusters with well-measured
deprojected gas density and temperature profiles available at present, it is clear that its size is
not large enough to draw statistically significant conclusions on the
viability of the effervescent heating model. Therefore, considering 
the simplicity of the procedure, any future work with the aim of
extending the analysis to a much larger sample will provide vital information
on this issue. 
\begin{acknowledgements}
We wish to thank the referee for insightful comments that improved
the presentation of the results. We also acknowledge Mateusz Ruszkowski 
for helpful remarks. This work is based on observations
obtained with XMM-Newton, an ESA science 
mission with instruments and contributions directly funded by 
ESA Member States and the USA (NASA). RP acknowledges support
from the Swiss National Science Foundation and the Tiroler
Wissenschaftsfond. SRON is supported financially by the NWO, 
the Netherlands Foundation for Scientific Research.
\end{acknowledgements}

\end{document}